\documentclass[twocolumn,prb,aps]{revtex4}
\usepackage{amsmath}
\usepackage{amssymb}
\usepackage{graphicx}
\usepackage{dcolumn}
\usepackage{graphicx}
\usepackage{dcolumn}
\usepackage{bm}

\begin{document}

\preprint{Phys.Rev.B}

\title{Mesoscopic transport in two-dimensional topological insulators}
\author{G. M. Gusev,$^1$  Z.D Kvon,$^{2}$ E.B.Olshanetsky,$^{2}$ and N. N. Mikhailov,$^2$  }

\affiliation{$^1$Instituto de F\'{\i}sica da Universidade de S\~ao
Paulo, 135960-170, S\~ao Paulo, SP, Brazil}
\affiliation{$^2$Institute of Semiconductor Physics, Novosibirsk
630090, Russia}

\date{\today}

\begin{abstract}
Topological states of matter have attracted a lot of attention due to their many intriguing transport properties. In particular, two-dimensional topological insulators (2D TI) possess gapless counter propagating conducting edge channels, with opposite spin, that are topologically protected from backscattering. Two basic features are supposed to confirm the existence of the ballistic edge channels in the submicrometer limit: the 4-terminal conductance is expected to be quantized at the universal value $2e^{2}/h$, and a nonlocal signal should appear due to a net current along the sample edge, carried by the helical states. On the other hand for longer channels the conductance has been found to deviate from the quantized value. This article reviewer the experimental and theoretical work related to the transport in two-dimensional topological insulators (2D-TI), based on HgTe quantum wells in zero magnetic field. We provide an overview of the basic mechanisms predicting a deviation from the quantized transport due to backscattering (accompanied by spin-flips) between the helical channels. We discuss the details of the model, which takes into account the edge and bulk contribution to the total current and reproduces the experimental results.

\end{abstract}

\maketitle

\section{Introduction}

The concept of a topological insulator (TI) has appeared in the condensed matter physics relatively recently, in 2007. Topological insulators are a new class of materials which are characterized by
 a bulk band gap like an ordinary band insulator, but have protected conducting states at their edge in the case of two-dimensional TI or of the surface for three-dimensional TI. The most
 important feature of TI is that its
behavior is independent of its specific geometry.

\begin{figure}
\centering
\includegraphics[width=9cm]{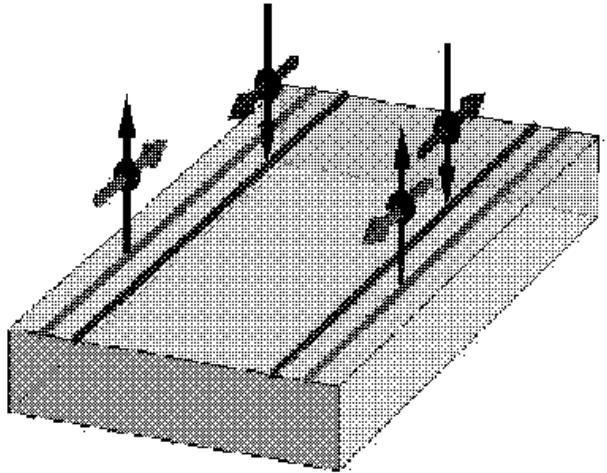}
\caption{Schematic drawing of the 2D topological insulator slab shaped sample with counter propagating spin polarized edge states. The left edge contains a forward mover with a spin up and a backward mover with a spin down. The spin and momentum direction are reversed for the right edge.}
\end{figure}

It has been shown that certain materials with a strong spin-orbit coupling (SOC) can demonstrate very intriguing phenomena. Mathematically, one can construct invariant integrals
over the momentum space which rely on the symmetry of electron wave function and result in the energy bands inversion. Particulary, such an inversion occurs due to the spin-orbit coupling
and the Darwin term
contributions to the hamiltonian of the crystals formed from heavy atoms.

SOC is a relativistic effect equivalent to an internal magnetic field without a violation of the time reversal (TI) symmetry. In two-dimensional materials with strong SOC there is a spin-up forward mover and a spin-down backward mover at the left edge, as illustrated in fig 1. At the right edge, the spin and the associated momentum directions are reversed, as is shown in the same figure. A system with such edge states is said to be in a quantum spin Hall (QSH) state, because it has a net spin current flowing forward along the top edge and backward along the bottom edge, just like the separated charge transport in the quantum Hall state. This phenomena
was predicted by Kane-Mele \cite{Kane} and Bernevig-Huges-Zhang \cite{Bernevig} in theoretical models considering spin orbit coupling. Although the edge states include both the forward and backward moving carriers, the backscattering between these states by a non-magnetic impurity is forbidden. The reason for this can be attributed to the fact that the edge state backscattering should necessarily involve a spin flip.
If the impurity carries a magnetic moment, then the time-reversal symmetry is broken and the edge state backscattering is possible due to the spin-flip process caused by the magnetic impurity.
In that sense the QSH edge state is protected against backscattering by the TR symmetry. The possibility of obtaining topologically protected, dissipationless spin current through 2D systems can
be very useful for future generation spintronic devices.

If the Fermi energy is tuned into the bulk energy gap of a conventional insulator, electron transport will be absent. On the other hand, when tuning the Fermi energy into the bulk energy gap of a topological insulator, electron transport becomes possible via the edge states. In general, the presence of impurities or defects in one-dimensional channels has a negative effect for the conductance, decreasing the elastic mean free path.
In 2D topological insulator backscattering is forbidden and ballistic transport is expected to be insensitive to nonmagnetic impurities and disorder. Important experimental consequences expected for the edge transport
 are the conductance quantization with the universal value $2e^{2}/h$ \cite{Konig}, and nonlocal voltage measured away from the dissipative bulk current path \cite{Roth}.

 In our review we focus on a number of issues that have been addressed in the experiments on mesoscopic HgTe samples and in the theoretical models that consider the mechanisms responsible for the experimentally
 observed deviations from the ideal TI. In Section 2 we provide the description of HgTe quantum wells, which have been used to study the transport in 2D topological insulators. In Section 3 we discuss the edge
 states transport
  using the Landauer-B\"{u}ttiker formalism and simple Kirchhoff's circuit rule. In Section 4 we describe experimental results involving conductance quantization in HgTe-based topological insulators. In Sec. 5 we
  review several important experiments on nonlocal transport measurements in 2D TIs. In Sec. 6 we discuss the edge states transport in more detail by considering the backscattering mechanisms at the
  edge in a nonballistic regime. In Section 7 we provide a more detailed quantitative description of the edge state transport. We calculate the nonlocal resistance in the presence of scattering between the edge
  states running
  along the same boundary and also across the bulk (via the bulk states). Finally, Sec. 8 we briefly summarize the main topics discussed in this review.

\section{2D topological insulator based on HgTe quantum wells}

The most important property of all topological insulators is the presence of delocalized surface or edge states. It is worth noting, that already quite a long time ago the existence of the surface states has been
addressed within the framework of Tamm-like or Shockley-like models by investigating how surface bands originate from discrete levels of atoms. But the first serious calculations appeared in the pioneer work \cite{Dyakonov}. The authors of \cite{Dyakonov} were the first to show that the presence of a spin-orbit interaction leads to the appearance of surface states, and, in particular, on the surface of the mercury
telluride and at the boundaries
of the quantum well created on its basis. Following this work came a few more articles in which this question was discussed in relation to the valence band. The result of this early work was summarized
in \cite{Gerchikov}, where the band spectrum of the mercury telluride quantum well and all its basic size quantization properties, including the interaction of bulk and surface states and their mutual
transformation, were
calculated using the exact Kane Hamiltonian. Along with the works already mentioned a special notice should be given to \cite{Pankratov}, where the possibility of surface bands of massless Dirac fermions at the
interface between semiconductors with an inverse and a normal energy spectrum was for the first time pointed out. However, all this work was not backed up by experiment, because at the time there was no technology for the fabrication of the described quantum wells. The surge in the research of topological isolators occurred later when the new theoretical ideas proposed in \cite{Kane, Bernevig} were almost immediately confirmed by
experiment. This work in some respect (concerning the existence of the edge states) repeated the conclusions of the earlier theories mentioned above, but what was more important for the subsequent boom in the
 area, it demonstrated that all of these states can be unified within a universal concept of topological order. A popular brand found almost immediately – the topological insulator \cite{Moore}, -
 also contributed to make that topic an object of a considerable interest.

Let us dwell in more detail on the concept of topological order. It consists in introducing the $ Z_{2}$ topological invariant, which is expressed through the integral over the boundary of the bulk Brillouin zone \cite{Kane} and, in fact, reflects unambiguous relationship existing between the volume and the surface. In the case of a normal insulator $ Z_{2} = 0 $, while for the topological insulator $ Z_{2} = 1 $. To put
it simply, $ Z_{2} $ equals the number of allowed bands on the surface. Generally speaking, similar topological approach was developed by analyzing the quantum Hall effect long before the subject of the topological insulators was raised \cite{Thouless,Levine}. No wonder, then, that a 2D system in the QHE regime is now cited as an example of a two-dimensional topological insulator. Mathematically it is possible to
construct $ Z_{2} $ invariants in various ways, but their physical meaning is uniquely related to the symmetry of the wave function, which changes radically as a result of the band spectrum inversion. Such
 an inversion is due,
in fact, to the relativistic terms in the Hamiltonian of a crystal consisting of heavy atoms, such as $ Hg $ or $ Bi $. The main terms are two in number: the more important is due to the spin-orbit interaction
 and the less important is associated with the Darwinian shift. There are three types of spectrum inversion: $s-p$, $p-p$ and $d-f$ \cite{Zhang}. A special place in this series belongs to mercury telluride, in which,
 as is well known, the simplest type of $ s-p $ inversion is realized, in which the hole-like band $ \Gamma 8 $ lies $ 0.35 $ eV above the electron-like band $ \Gamma 6 $. However, despite the spectrum inversion,
 the 3D $ HgTe $ is not a topological insulator, since in its bulk a gapless state is realized, which can be altered only by lowering the initial symmetry of the crystal by some external influence. Such an external influence may, for example, be uniaxial compression \cite{Dai}. A special situation is realized in quantum wells based on HgTe. In such wells, as a result of dimensional quantization, at the well width above the
 critical, $ d_{c} = 6.3 $ nm - $ 6.5 $ nm, an inverse band gap appears in the two-dimensional bulk spectrum together with the gapless edge states at the well boundaries and, thus, a two-dimensional TI is realized,
 with which we will begin the main part of this review.

\begin{figure}
\includegraphics[width=9cm]{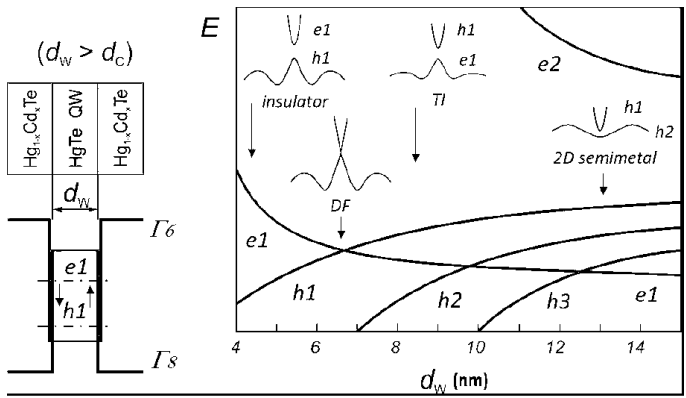}
\caption{
Schematic of the HgTe quantum well energy spectrum
 depending on the well thickness.}

\end{figure}

To begin, let us describe in more detail the energy spectrum of the quantum well based on mercury telluride. Figure 2 shows qualitatively dependence of the main size-quantized subbands extrema energy
on the well thickness. The spectrum modification with well thickness can be arbitrarily divided into three distinct regions: the region at $ d < d_{c} $,
where a direct band gap two-dimensional insulator is realized, the gap decreasing with the well thickness increasing. At the critical well thickness $ d_{c}$ equal, depending on the surface orientation and the
quantum well deformation, to $6.3 - 6.5 $ nm, the band gap collapses, and then with the width increasing further, a second region begins where the two-dimensional topological insulator with an inverted energy spectrum exists.
Finally,
for $ d > 15-16 $ nm a semi-metal state is realized due to the overlapping of the hole-like band $ h1 $ (the conductance band) and $ h2 $ (the valence band). Since further we will talk about the properties
of the two-dimensional topological insulators (DTI), we are interested only in the second region with the two-dimensional TI. The energy spectrum of this TI, calculated in \cite{Raichev}
for surface orientations (100)
and (013), is shown in Fig.3a. As one can see, the main characteristics of the spectrum depend only weakly on the surface orientation. In both cases the critical thickness $ d_{c} = (6.2 - 6.3) $ nm,
and the two-dimensional TI state with the largest band gap, corresponding to a simpler $ s-p $ type of inversion, is realized at a well thickness of $ (8.2-8.5) $ nm. In this case, the value of the gap is
  approximately $ 30 $ meV. The energy dispersion for the edge and the bulk states of a (013) $ 8.5 $ nm well is shown in Fig.3, b. The figure illustrates well all the features of the spectrum of a
  two-dimensional TI on the basis of $ HgTe $ quantum well: a linear Dirac spectrum of the edge current states and a parabolic gapped spectrum of the bulk states. Note that the edge states exist not
  only in the gap, but also at the energies corresponding to the allowed bulk energy bands. In Fig.3, b one can clearly see the anticrossings of the edge states branches in the lower part of bulk
  gap caused by a lower surface symmetry (013).

\begin{figure}
\includegraphics[width=9cm]{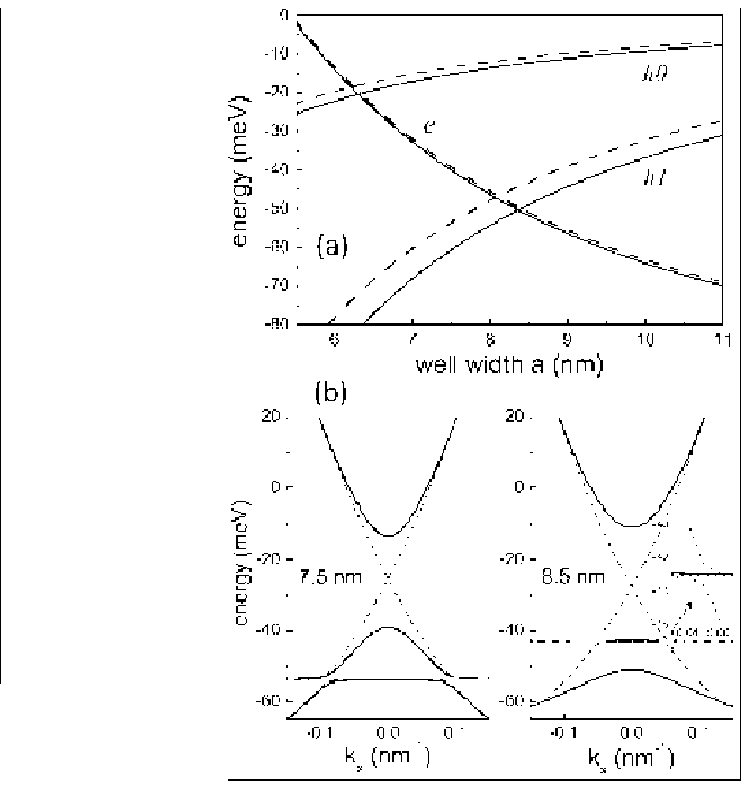}
\caption{
Spectrum near the critical thickness, b) the
energy dispersion for the bulk and the edge states.}
\end{figure}

\section{Landauer-B\"{u}ttiker formalism and Kirchhoff's circuit rule}
\label{3}

Before we start discussing experimental results, let us consider simple theoretical models, describing the edge state transport in 2D TI. To analyze the multiprobe resistance in the quantum Hall effect (QHE)
regime, B\"{u}ttiker developed the concept of chiral ballistic edge states. Since such state propagates only in one direction, it carries with it the same electrochemical potential, and the resistance
between two
neighbouring potential contacts along the edge is found to be equal to zero. The origin of resistance in 2D TI, like in the conventional quantum Hall effect, is due to the processes in the contact regions.
The contacts are assumed to be thermal reservoirs, where the mixing of electron states with different spins will occur. Note that, in contrast to the QHE, where the mixing of the edge states occurs within
metallic Ohmic contacts, in our samples its takes place in the 2D electron gas regions outside of the metallic gate due to a finite bulk conductivity and fast spin relaxation.

\begin{figure}
\includegraphics[width=9cm]{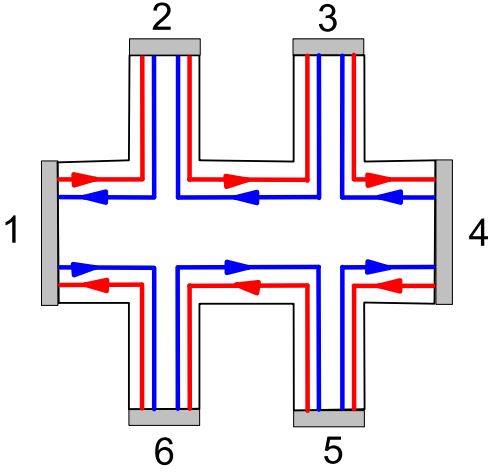}
\caption{(Color online)
Schematic of sample and counterpropagating spin-polarized edge states.}

\end{figure}

In 2D topological insulators the edge states propagate in two opposite directions. Indeed it is easy to generalize Landauer-B\"{u}ttiker formalism to counter-propagating states and to a multi terminal scheme. Let us start with a 6-probe Hall bar, used in the majority of conventional experimental set-ups, figure 4. The resistivities and conductivities follow from a set of equations \cite{Buttiker}:
\begin{equation}
I _ { i } = \sum _ { j } \left( G _ { j i } V _ { i } - G _ { i j } V _ { j } \right) = \frac { e ^ { 2 } } { h } \sum _ { j } \left( T _ { j i } V _ { i } - T _ { i j } V _ { j } \right)
\end{equation}
where $V_{i}$ is the voltage at {\emph{i}th terminal and $I_{i}$ is the current flowing from the same terminal. Here $T_{ij}$ is the transmission from the \emph{j}th to the \emph{i}th terminal and $G_{ij}$ is the corresponding conductance. The transmission coefficient in the absence of backscattering between edge states can be written in the form:
\begin{equation}
T = \left( \begin{array} { c c c c c c } { - 2 } & { 1 } & { 0 } & { 0 } & { 0 } & { 1 } \\ { 1 } & { - 2 } & { 1 } & { 0 } & { 0 } & { 0 } \\ { 0 } & { 1 } & { - 2 } & { 1 } & { 0 } & { 0 } \\ { 0 } & { 0 } & { 1 } & { - 2 } & { 1 } & { 0 } \\ { 0 } & { 0 } & { 0 } & { 1 } & { - 2 } & { 1 } \\ { 1 } & { 0 } & { 0 } & { 0 } & { 1 } & { - 2 } \end{array} \right)
\end{equation}
Consequently we can set $I_{2} = I_{3} = I_{5} = I_{6} = 0$ and it is easy to obtain
\begin{equation}
\begin{array} { l } { G _ { 4 t } = \frac { I _ { 14 } } { \mu _ { 3 } - \mu _ { 2 } } = \frac { 2 e ^ { 2 } } { h } } \\ { G _ { 2 t } = \frac { I _ { 14 } } { \mu _ { 4 } - \mu _ { 1 } } = \frac { 2 } { 3 } \frac { e ^ { 2 } } { h } } \end{array}
\end{equation}
where $G _ { 4 t }$ and $G _ { 4 t }$ are 4-terminal and 2-terminal conductances respectively. Remarkably, the 2D topological insulators are characterized by nonlocal effects: the application of a current between any
pair of the probes creates a net current along the sample edge which can be detected from any other pair of voltage probes. One can say that there is no difference between local and nonlocal electrical measurements
in a topological insulator. For example, for the sample shown in Fig.4, the application of current between leads 2 and 6 produces current along two paths: a longer one , 2–3, 3–4, 4–5, 5–6, and a shorter one, 2–1, 1-6.
The non local signal in this geometry will be measured between the probes 3-5. The Landauer-B\"{u}ttiker formalism allows to derive the nonlocal conductance for any measurement configuration in the 6-probe Hall bar,
shown in figure 1. Moreover, using the universal approach, the nonlocal response can be found for ballistic transport in an arbitrary N-terminal sample \cite{Protogenov}. Actually, to find the value of the local and nonlocal resistance in most of the experimental set-ups a simple picture of helical edge states is usually enough. Using Kirchhoff’s rules, the 1D channel between any contacts can be substituted by the quantum resistance $R=h/e^{2}$. In this case a simple expression can be derived that allows one to calculate the resistance value for any measurement configuration assuming that there is only diffusive edge state transport in
the sample \cite{Olshanetsky}:
\begin{equation}
R_{n,m}^{i,j}=\frac{L_{n,m}L_{i,j}}{Ll} (h/e^2)
\end{equation}
where $R_{n,m}^{i,j}$ is the voltage measured between contacts $i$ and $j$ while the current is maintained between contacts $n$ and
$m$, $L_{i,j}$ ($L_{n,m}$) are the distances between $i$ and $j$ ($n$ and $m$) along the gated sample edge that does not include $n$ and
$m$ ($i$ and $j$), $L$ is the total perimeter of the sample, and $l$ is the mean free path due to the scattering between helical states
propagated along the same edge.

\section{Conductance quantization : Experiment}
\label{4}

\begin{figure}
\includegraphics[width=9cm]{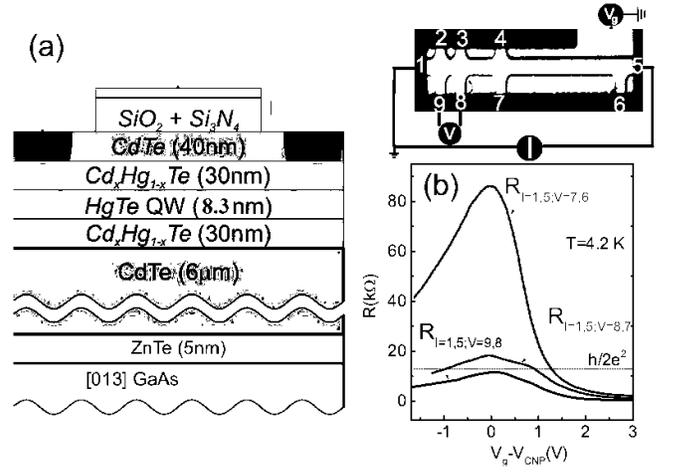}
\caption{
(a) Schematic of the transistor and the top view of the sample. (b) Resistance R as a function of gate voltage measured between various voltage probes, T=4.2 K. }
\end{figure}

\begin{figure}
\includegraphics[width=9cm]{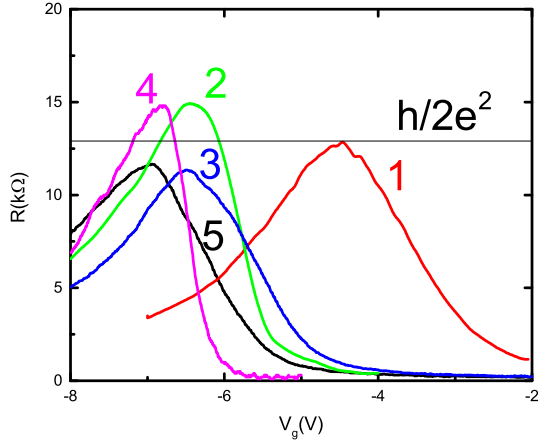}
\caption{\label{fig.6}(Color online)
Resistance R as a function of gate voltage measured for different samples, T=4.2 K.}
\end{figure}

\begin{table}[ht]
\caption{\label{tab1} Typical parameters of the electron system in HgTe quantum well at T=4.2K.}
\begin{ruledtabular}
\begin{tabular}{lccccc}
sample & $d$ (nm) & $V_{CNP}$ (V)& $R_{max}(h/2e^{2})$& $\mu (V/cm^{2}s )$ \\
1& 8.3 & -4.5& 1.0 & 12.800\\
2& 8.4 & -6.4 & 1.15 &62.500\\
3& 8.3 & -6.5 & 0.88 &25.600 \\
4& 8.3 & -6.83& 1.15 &120.200 \\
5& 8.3 & -6.95& 0.9 &34.400 \\
\end{tabular}
\end{ruledtabular}
\end{table}

\begin{figure}
\includegraphics[width=9cm]{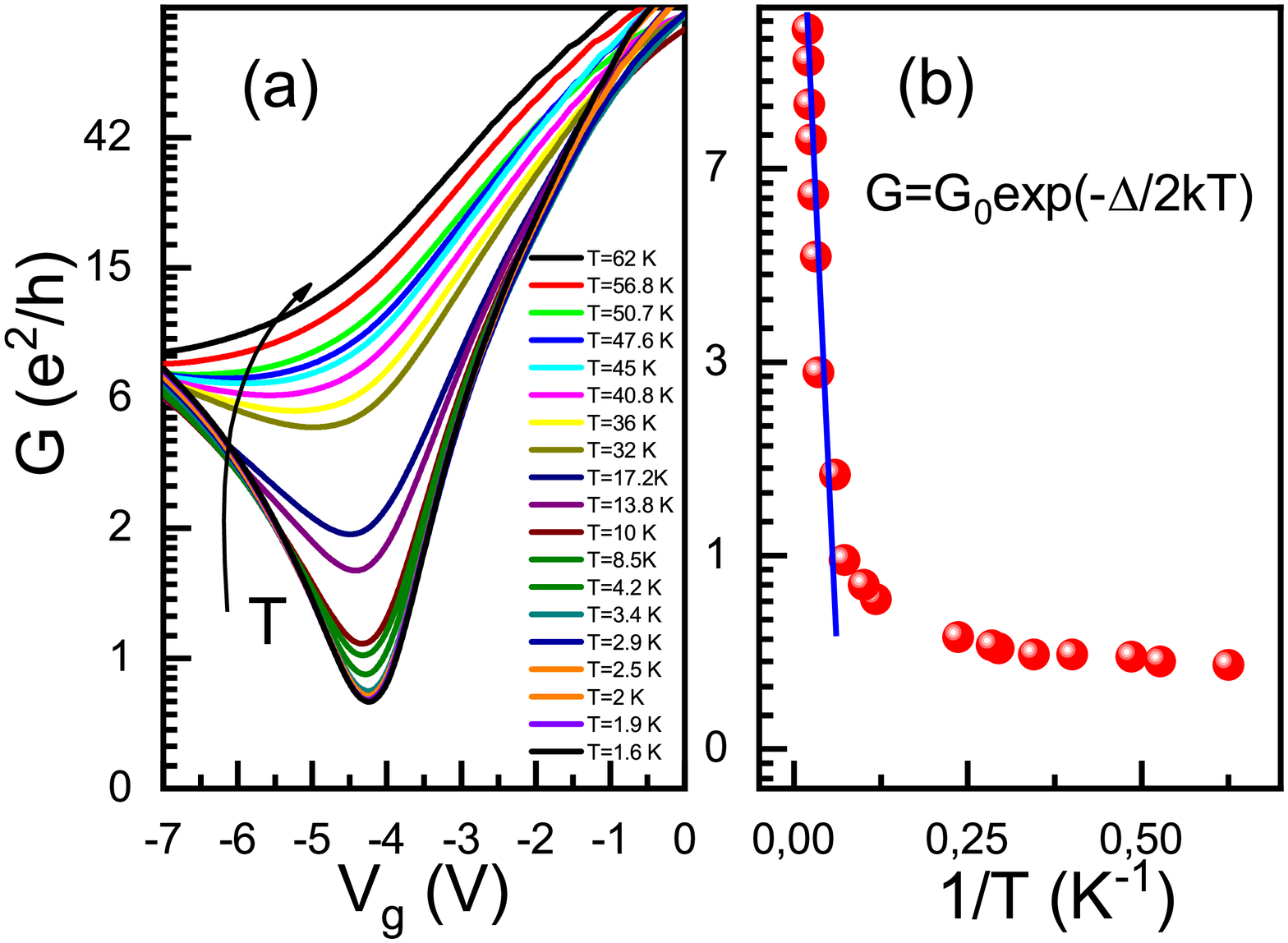}
\caption{(Color online)
(a) Conductance of the shortest segment (I=1,5;V=9,8) as a function of the gate voltage for different temperatures. (b)Conductance at CNP
as a function of 1/T. }
\end{figure}

The experimental samples used to obtain the data discussed in this review were practically all made on the basis of quantum wells with the thickness $ 8 $ or $ 8.3 $ nm and with the surface orientation (013). The choice of this surface orientation is on the one hand due to the fact that the presence of steps on such surface provides a more balanced growth of the $ HgTe $ and $ HgCdTe $ layers, thus respectively reducing the concentration of various kinds of point and dislocation defects and, on the other, to the fact that, as shown in Fig.3, the energy spectrum of a two-dimensional TI is practically independent of the surface orientation. It is also important to note that as far as the thickness of the quantum well is concerned, its exact value for each given sample may not match the specified growth thickness and deviations from it by a few tenths of a nanometer due to some inhomogeneity of the atomic beam density in MBE growth are quite possible.

To study the edge states properties of a two-dimensional TI in a HgTe quantum well two conditions are necessary: the Fermi level must be located in the bulk band gap and one has to have a clear and convincing way to detect the edge states. The first condition may be fulfilled by using a field effect transistor structure whose schematic image is shown in Fig.5a ( on the right-top one can see the top view of the sample). The second condition will be discussed in detail in the next section. The manufacture of the field effect transistors based on the $ HgTe $ quantum well requires two more operations: low-temperature growth of the dielectric layer, and the fabrication of a metal gate on top of it. As a dielectric either the pyrolytic layer $ SiO_{2} $ or the double layer $ SiO_{2} + Si_{3}N_{4} $ grown at temperatures of $ 80-100^{\circ} $ C is usually employed. The gate is $ Ti / Au $ layer. Note that there are other methods of growing dielectric layers which are not discussed here. The density variation with gate voltage is $(1.09\pm 0.01)\times 10^{15} m^{-2}V^{-1}$.
The transport measurements in the investigated two-dimensional TI were conducted in the temperature range of $ 0.2-70 $ K using a standard phase-sensitive detection scheme at frequencies $ 2 - $ 12 Hz
and at driving current levels $ 0.01 - $ 10 nA to avoid the heating of the electron system. Figure 5 (b) shows resistance as a function of the gate voltage measured between different probes at T=4.2 K. It is worth noting that the edge current flows along the gated sample edge whose length $L_{gate}$ is longer than the distance between the probes L (bulk current path) and corresponds to $5-6 \mu m$. For longer distances between the probes
we see higher resistances. The resistance is low (about 100 Ohm) at the gate voltages, corresponding to the position of the Fermi level ($ E_{F} $) in the conduction band, passes through a maximum (in this case approximately equal to $ 12.9 k\Omega$ for the shortest distance between probes), corresponding to $ E_{F}$ going through the middle of the bulk bands gap, and then begins to decrease, reaching values of few $k\Omega$ when the Fermi level enters the valence zone. The voltage (or rather the Fermi level position) corresponding to the $ R $ maximum is called the charge neutrality point (CNP). We have also measured the Hall effect (Hall resistance $ R_{xy} (V_{g}) $) in this samples (not shown). The $ R_{xy} (V_{g}) $ dependence shows distinct plateaus with $ i = 1 $ and $ i = 2 $ on the electron side, at the CNP it goes through zero, and, with the Fermi level in the valence band, it changes sign but plateaus are no longer observed due to a significantly (by an order of magnitude) lower hole mobility. The absence of the Hall signal at the charge neutral point indicates that there are no mobile charge carriers in the quantum well. Note that, strictly speaking, a zero Hall signal, and, moreover, the resistance maximum are not a direct evidence of the absence of charge carriers in the well and therefore each time a care should be taken when analyzing the CNP.

Figure 6 shows the resistance $R(V_{g})$ at zero magnetic field for 5 samples with different well thickness fabricated from different wafers. The table 1 lists the devices and indicates typical parameters, such as the well width d, gate voltage corresponding to the charge neutrality point position $V_{CNP}$, the value of the resistivity at CNP $\rho_{max}$ and the electron mobility $\mu = 1/(N_{s}e\rho) $ at the density $N_{s} = 2\times10^{11} cm^{-2}$. The position of the peak is usually related to the charge trap in the oxide, and it would be expected, that the resistance peak increases and its width becomes wider with a shift of $V_{CNP}$ . Surprisingly, we do
 not observe such effect. On the contrary, neither the peak value, nor the width are sensitive to the sample electrostatics. This observation proves that the transport properties at the CNP are due to the intrinsic properties
of the 2D TI edge states. The variation of the conductance with gate voltage and lattice (bath) temperature is shown in Figure 7a. The conductance between short probes reveals a broad minimum whose value is $0.5e^{2}/h$ , that is lower that is expected for the ballistic case. We see that the conductance increases sharply for temperatures above 15 K while saturating below 10 K. We find that the profile of the conductance temperature dependencies above $T > 15 K$ fits very well the activation law $G \sim exp(-\Delta/2 kT)$, where $\Delta$ is the activation gap. Figure 7b shows the evolution of the conductance-voltage profile with temperature.
 The thermally activated behavior of conductance above 15 K corresponds to a gap of 15 meV between the conduction and valence bands in the HgTe well. The mobility gap can be smaller than the energy gap due to
 disorder.
Below $ T < 10 K$ the conductance is saturated with the temperature, demonstrating no significant temperature dependence.

\section{Nonlocality:Experiment}

In general, dependencies like that shown in Figs.5-6 are not very informative about the edge transport, since such measurements do not provide for the elimination of the bulk contribution. The major approach to a straightforward detection of the edge states is a measurement in a non-local geometry.

\begin{figure}
\includegraphics[width=9cm]{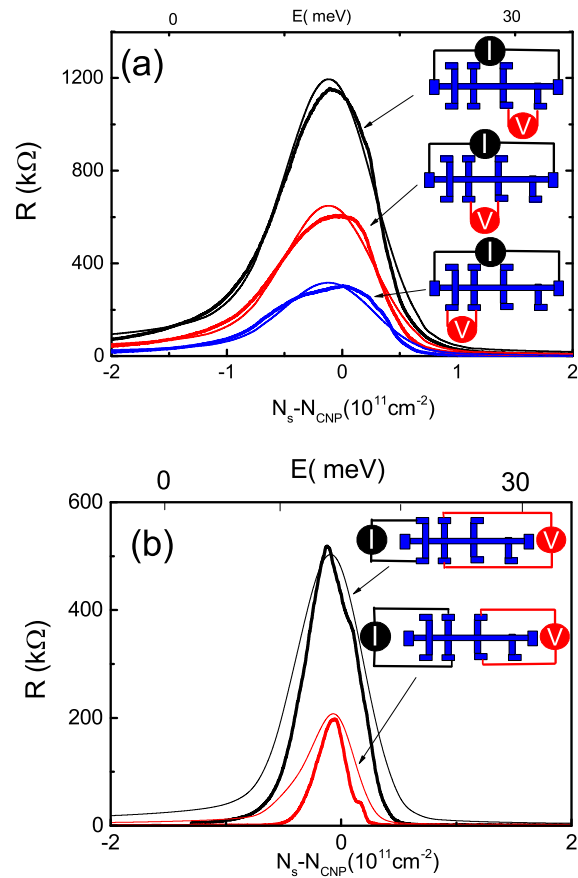}
\caption{(Color online) (a) Comparison of local resistances for three segments of the HgTe sample. (b) Nonlocal resistances for two different measurement configurations of the sample. The thin traces correspond to the prediction of the "edge+bulk" transport model. }
\end{figure}

Consider now a two-dimensional conductor with the resistivity $\rho_{xx}$, the length $ L $, the width $ W $ and the contacts 1-6, as shown in Fig.4. Then, if the current is passed through contacts 2,6 and the voltage is measured between contacts 3,5, the resistance $ R_{2,6}^{3,5}= V_{3,5} / I_{2,6} $ will be of the following order of magnitude $ R_{2,6}^{3,5} \approx \rho_{xx}\exp (- \pi L/W)$. That is, it will fall exponentially with
the length to width ratio of the conductor. The reason is trivial - only exponentially small part of the total current will reach the contacts 3,5. This is a configuration corresponding to a non-local resistance measurement. Now suppose that the bulk spectrum of the sample in question has a gap with the Fermi level lying in the middle of it. In the case of an ordinary insulator there will be no current. However, if the insulator is topological, then all the current will be carried by the edge states, since they are delocalized. Then, in the case of ballistic transport, we can apply Landauer-B\"{u}ttiker formalism or Kirchhoff's circuit rule, as has been shown in the section 3. From eq.4 we get for equal distances between all probes:$R_{2,6}^{3,5}=\frac{2}{3} (h/e^2)$. Thus, a comparative analysis of the local and non-local response allows one to unequivocally determine the presence of edge transport and, accordingly, of the current carrying edge states in the sample. A comparative analysis for ballistic 2D TI has been performed in 6 and 4-probe devices, and the results were successfully explained by the Landauer–Büttiker model \cite{Roth}. The case of a diffusive disordered 2D TI has been consider in \cite{Olshanetsky, Rahim}. As an example, Figure 8 shows typical measurement results for the local ($ R_{loc} $, Fig.8a) and non-local ($ R_{nonloc} $, Fig.8b) resistance in a nonballistic sample, based on $ 8 $ nm $ HgTe $ quantum well with the configurations of the current and voltage probes presented in the insert to this figure.
At a first glance, the behavior of these resistances is qualitatively similar and coincides with the dependence $ R(Vg) $ discussed above, Figs.5,6. However, a closer look reveals a significant difference in the behavior
of $ R_{loc} $ and $ R_{nonloc} $: while $ R_ {loc} $ has a noticeable value at all gate voltages, including those that correspond to the Fermi level position in the allowed bulk bands, $ R_ {nonloc} $ is close to zero
 at the indicated voltages as, indeed, it should be, since the edge states are short-circuited by the bulk states. However, it becomes comparable to $ R_ {loc} $ in the vicinity of the CNP, that is, when the Fermi level
 is located in the center of the bulk gap. It is such features of the transport in a two-dimensional TI that are indicative of the presence of the charge transfer along its edge. The first experiments on the edge transport were carried out in \cite{Konig,Roth}, where the ballistic edge transport was demonstrated in samples of submicron size based on $ 7-8 $ nm HgTe quantum wells. Then, in \cite{Gusev, Gusev2, Grabecky}, it was shown that
 the transport via the edge states persists in these quantum wells on a macroscopic scale of about one millimeter, but already in the diffusive regime. Finally, let us mention the experiments \cite{Konig2, Katja} on the visualization of the edge states, confirming their presence. However, a detailed discussion of these experiments goes beyond the scope of this review.

 \section{Deviation from conductance quantization: models and comparison with experiment in HgTe well}

As indicated in the previous chapter, when the Fermi level lies in the bulk gap of a 2D TI, one would expect that in small samples at sufficiently low temperatures the edge states will play an important or even a dominant role in the transport, resulting in the quantization of the resistance with the universal value $h/2e^{2}$, and in nonlocal resistance of the order of $\sim h/e^{2}$ in the absence of the bulk contribution to the transport. A number of the experimental works indeed confirm the presence of nonlocal transport \cite{Roth,Gusev,Olshanetsky}, but the longest edge channel length that shows a quantized resistance plateau never exceeds $10 \mu m$ \cite{Olshanetsky}. The lack of robustness of the edge state transport to disorder attracted a lot of attention, and a large number of various explanations have been proposed in recent years. We review several of the most prominent theoretical models of the helical edge states scattering and compare them with the experiment. It is worth noting that in the presence of electron-electron interactions, the edge states of 2D topological insulator can be regarded as a helical Luttinger liquid (LL) \cite{xu,wu,teo,voit,kane2,gorny,schmidt} with important consequences for the scattering mechanisms discussed below.

The most effective way to induce spin flip scattering is a magnetic impurity. It has been predicted \cite{maciejko} that at high temperature the backscattering by magnetic impurity leads to a small negative correction to the conductance. With decrease of temperature, the Kondo effect enhances the backscattering and the correction grows. The rate of the spin-flip process depends on the interaction parameter of the single nonchiral Tomonaga-Luttinger liquid \cite{Kane, gorny, Kainaris}.

Note, however, it is highly unlikely that a magnetic impurity should be present in the HgTe and CdTe material grown by molecular beam epitaxy \cite{mikhailov}. Scattering from residual disorder comes from several origins: remote charged impurity scattering from Si donors, alloy disorder scattering, interface roughness scattering, and uniformly distributed background charged nonmagnetic impurities are a few examples.

Many models have been proposed as an explanation of the backscattering at the edge of 2D TI due to nonmagnetic impurities or other mechanism, such as the effects of Rashba spin-orbit coupling \cite{strom, crepin}, phonons \cite{budich}, nuclear spins \cite{hsu}, noise \cite{vayrynen} and disordered probes \cite{mani}.

\begin{figure}
\includegraphics[width=9cm]{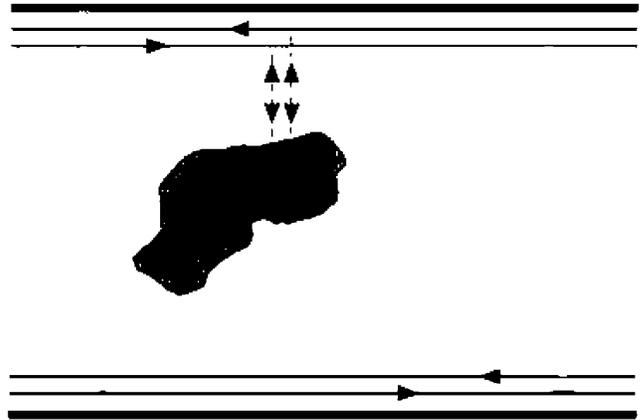}
\caption{
The counter propagating edge states tunnel-coupled to a conducting puddle in the bulk of the 2D topological insulator. The electron spin in the edge states is locked to the direction of motion, but the electrons in the puddle can flip the spin due to inelastic scattering (small puddle)\cite{vayrynen2} or strong spin-orbit interaction (large puddle) \cite{aseev}.}
\end{figure}

It is important that in such narrow band gap materials as HgTe potential fluctuations play a significant role. For example, potential fluctuations due to a nonuniform doping lead to formation of conducting puddles in the bulk of the insulator, and electrons at the edge states interact with these puddles, as shown in figure 9 \cite{vayrynen2}. The authors of the model \cite{vayrynen2} suggested that the puddles should be small and rare in order to provide small tunneling probability to the bulk, while, on the other hand, a number of puddles should be present in the vicinity of the edge to allow for spin flip between the counter-propagating states. Self-averaging resistance of a sample with a dominant edge state contribution to transport is given by \cite{vayrynen2}:
\begin{equation}
R\sim \frac{h}{e^{2}}\frac{1}{g^{2}}n_{p}\lambda \left(\frac{T}{\delta}\right)^{3}L
\label{eq5}
\end{equation}

where $n_{p}$ is the density of the puddles, $\lambda=\hbar v/E_{g}\approx 18 nm$ is the electron penetration depth into the puddles ($v\approx 5.5\times10^{7} cm/s$ is the electron velocity, $E_{g}\simeq 20 meV$ is the forbidden gap), $g$ is the dimensionless conductance within the dot (puddle), $\delta$ is the mean level spacing within the dot, L is the distance between probes (length of the edge states). Rewriting the eq.5 in the form $R=\rho_{0}L$, we obtain $\rho_{0}=15\times10^{3}(h/e^{2})/cm = 1.5(h/e^{2})/\mu m$. This confirms that the coherent ballistic transport might occur on the micron length scale. The density of the puddles can only be roughly estimated from the ratio of the total carrier density to the average number of electrons in the puddles. Note, that the puddles become populated when the local potential fluctuations exceed half of the band gap. The resulting equations for the characteristic donor density $n_{0}$ and the density of puddles $n_{p}$ have been obtained in \cite{vayrynen2}:
\begin{eqnarray}
n_{0}=\frac{E_{g}^{2}\kappa^{2}}{8\pi e^{4}ln\{l_{g}^{2}/[2l_{g}-l_{d})l_{d}]\}},\\
n_{p}\sim \left(\frac{1}{l_{g} a_{B}}\right)\left(\frac{n_{d}}{n_{0}}\right)^{1/2} \exp(-n_{0}/n_{d}),
\label{eq2}
\end{eqnarray}

where $\kappa=13$ is the dielectric constant, $l_{g} \approx 343 nm$ is the distance to the gate, $l_{d}\approx 8 nm$ is the distance to the donors, $n_{d}\sim 2\times 10^{11} cm^{-2}$ is the donor density, $a_{B}=\frac{2\hbar v}{\alpha E_{g}}\approx 120 nm$ ($\alpha=e^{2}/(\kappa hv)=0.3$). From equations 6 and 7 we find $n_{0}\approx 4\times10^{10} cm^{-2}<< n_{d}$ and $n_{p}\approx4\times10^{9} cm^{-2}$. The dimensionless parameter $g$ can be estimated as $g\sim\sqrt{N}\approx 1-2$, where N is the number of the electrons in the puddle $N\sim a_{B} n_{d}^{1/2}\approx 2-5$. Combining all parameters we finally calculate $\rho_{0}=(2-8)\times10^{3}\left(\frac{T}{\delta}\right)^{3}(h/e^{2})/cm$. The energy level spacing is estimated from Coulomb blockade energy in the dot $\delta \sim \alpha^{2} E_{g}\sim 1-2 meV$. At relatively high temperatures $T\approx 10 K$ we obtain $T \sim \delta$ and $\left(\frac{T}{\delta}\right)^{3}\sim 1$ and, it is expected, that the T dependence is saturated. In this case calculations give a result comparable with the experimental value.

An important experimental observation, that can discriminate between different theoretical models could be the temperature dependence measurement of the conductance at the charge neutrality point, where the edge transport contribution is predominant. The experiments \cite{Gusev2} show a nearly temperature-independent conductance in disordered 2DTIs.
However, quasiballistic HgTe samples demonstrate a weak linear T dependence. An example of such behaviour is shown in Fig.10a.

Note, that we don`t see any $T^{3}$ dependence predicted by the model \cite{vayrynen2}, in a wide temperature interval below 10 K (figure 7). Large puddles would yield a large parameter g resulting in a small value of the relative resistivity $\rho_{0}$. Therefore, our attempts to account for a weak temperature dependence only increase the discrepancy between theory and experiment. However, if a large enough number of puddles are situated immediately at the edge of the sample interrupting the edge current states flow, perhaps the resulting temperature dependence would be closer to that observed in the experiment.

One can suppose that the size of the puddles is large enough to allow for a two-dimensional electron states with continuous energy spectrum, \cite{essert, aseev}. The impurity scattering in the puddles combined with spin-orbit coupling may result in a temperature-independent spin relaxation due to spin orbit interaction. The existence of large-size puddles could lead to an effective backscattering of electrons. For example, the authors of the model \cite{aseev} demonstrate, that one puddle reduces the conductance by half if the tunnel coupling and spin-flip scattering in the puddle are sufficiently strong. Note, however, that a comparison with experiment requires an accurate and a more quantitative description of the spin orbit relaxation mechanism which is not yet available.

Another point of view is that the suppression of the conductance quantization does not result from the TRS violation by scattering mechanisms but rather relates to the properties of the sample edge potential \cite{wang}. Indeed the notion of the 2D TI with helical edge states protected against backscattering by the time reversal symmetry is only valid for sharp boundary conditions. The realistic smooth edge potential leads to the edge reconstruction, and consequently to spin separation \cite{wang}. It has been shown that when the degree of smoothing or the effective edge width increases, the TRS starts to be spontaneously broken, and elastic single-particle backscattering is allowed. In particular, the authors \cite{wang} found that for appropriate parameters of the HgTe quantum well, an edge-reconstruction transition occurs, and the bulk electron density drops to zero at the edge on a scale of 10 nm. Interestingly, the model predicts a possible transmission blockade through two quantum point contacts in series fabricated on top of 2D TI in the edge transport regime \cite{wang}. Still numerous technological challenges need to be addressed for fabrication of such device.

Interaction plays an important role in yet another model recently proposed for 2D TI \cite{novelli} . Nonmagnetic short range impurity surrounded by interacting electrons forms a local magnetic moment. This scenario is valid for atomically thin crystals, like $WTe_{2}$, where vacancy defect naturally occur at the edge, \cite{wu}, but is unlikely in HgTe wells, where the puddles scenario due to smooth potential fluctuations is expected \cite{vayrynen2, essert, aseev}.

While the topological protection forbids elastic backscattering from nonmagnetic impurities, no such simple result exists for inelastic backscattering when interactions are present. In \cite{Kainaris} the authors have studied the transport properties of a generic one-dimensional helical liquid in the presence of interactions and disorder. They considered three dominant scattering mechanisms. The first mechanism, the inelastic single particle scattering mechanism, denominated as a 1P process, leads to a change in the chirality of a single incoming particle. The second important mechanism describes the inelastic backscattering of two electrons -2P process (see figure 10b). Finally, the authors argue that the backscattering may occur due to electron-electron interactions only \cite{Kainaris}. The main results are summarized as follows:

 \begin{equation}
\Delta G \sim \frac{e^{2}}{h}L_{edge} T^{-2K-2}, K > 2/3
\end{equation}
 \begin{equation}
\Delta G \sim \frac{e^{2}}{h}L_{edge} T^{-8K+2}, K < 2/3
\end{equation}

where $K$ is the Luttinger liquid parameter. It is important to note that, at $K > 2/3$, transport properties are dominated by 1P scattering, and below $K = 2/3$, the 2P process becomes important. In addition it is expected that, below $K < 3/8$, the localization of the helical edge states takes place. Assuming the Luttinger parameter $K\approx3/8$ (2P process), which corresponds to the weak coupling regime, we obtain a good agreement with the experimental dependence $\Delta G \sim \frac{e^{2}}{h}L T^{-1}$, as one can see in figure 10a (remember, that $G=2e^{2}/h-\Delta G$ in our case). Intriguingly, this parameter value also marks the transition from the localized to the delocalized regime in the transport of the disordered helical liquid.
\begin{figure}
\includegraphics[width=9cm]{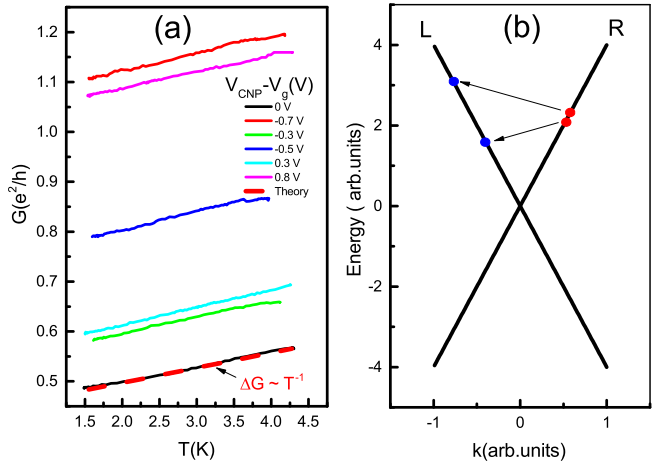}
\caption{(Color online) (a) The conductance as a function of the temperature for different gate voltages. Dashes-linear dependence for conductance corrections predicted by the theory \cite{Kainaris}. (b) Schematic of the inelastic scattering process of two electrons described by the model \cite{Kainaris}, referred as "2P process" and describing
the scattering of the interacting particles in disordered helical Luttinger liquid. }
\end{figure}

 It is worth noting that a previous study of strongly disordered topological insulators revealed a practically temperature independent resistance \cite{Gusev2} with, however, some tendency to localization behaviour. It is also worth noting that the electron-phonon scattering results in a $T^{3}$ dependence of the scattering rate, which disagrees with our observations.

 To conclude this section we expect that the formation of local large size puddles due to impurity potential fluctuations is a general feature of 2D topological insulators based on narrow gap semiconductors, such as HgTe quantum wells. The backscattering may arise from the tunnel coupling between the edge states and conducting puddles. Alternatively, consideration of one-dimensional helical liquid in the presence of interactions and disorder can explain the weak temperature dependence of the conductivity near the CNP, which has been observed in HgTe samples. Discrimination between different models requires further theoretical and experimental study.

\section{Deviation from conductance quantization:the edge $+$ bulk model }

Landauer-B\"{u}ttiker formalism predicts the values of the local and nonlocal resistancies for any N-terminal configuration scheme in a ballistic device. Kirchhoff's circuit rule allows to calculate N-therminal resistance assuming that there is only edge state transport along the perimeter of the sample, in the presence of backscattering between channels near the same edge. Indeed both models describe the transport at the CNP and neglect the bulk contribution. In this chapter we provide the model which takes into account the edge and bulk contribution to the total current and reproduces our experimental results.

Figures 5,6 and 8 demonstrate that when the gate voltage variation causes the Fermi level to move from the bulk electron states to the bulk hole states via the gap with helical edge states the resistance of a HgTe quantum well shows a broadened peak around the charge neutrality point. The character of the conductivity depends on the position of the Fermi level. When the Fermi energy lies in the conduction or valence bands, the edge states coexist with the bulk states and the mixing between the boundary and the bulk may lead to a strong backscattering. Figure 11 shows schematically edge state propagations and dominant scattering process, when the Fermi level moves from the valence to the conductance bands. We denote $\varphi_{i}$, and $\varphi_{i^{\prime }}$, as the potentials at the opposite (bottom and top, respectively) edges of the sample. Indexes $i=1,2$ label the states with different projections of the spin. The scattering between the edge states and the bulk can be described by a phenomenological parameter $g$, which is related to the edge-bulk scattering rate, and backscattering along the border can be described by a single phenomenological parameter $\gamma$, which is related to the edge–edge scattering rate. One can see that the scattering between the edge and the bulk becomes important, when the Fermi level lies in the conduction or in the valance bands ( Fig.11a,c), and can be negligible, when the Fermi level crosses the gap (Fig.11b).

It is important to consider the density of states in order to determine the electron and hole bulk densities and bulk conductivity. Figure 12a shows schematically a profile of the density of states in a HgTe quantum wells. Because of the random potential the conduction and valence bands have Gaussian tails stretching into the band gap. According to the generally accepted theory the electrons and holes in the band tails should be localized. However for simplicity we assume a finite residual conductivity in the band tails in order to explain the reduction of nonlocal transport near the CNP. The insert to fig 12a shows schematically energy spectrum of topological insulator.

\begin{figure}
\includegraphics[width=9cm]{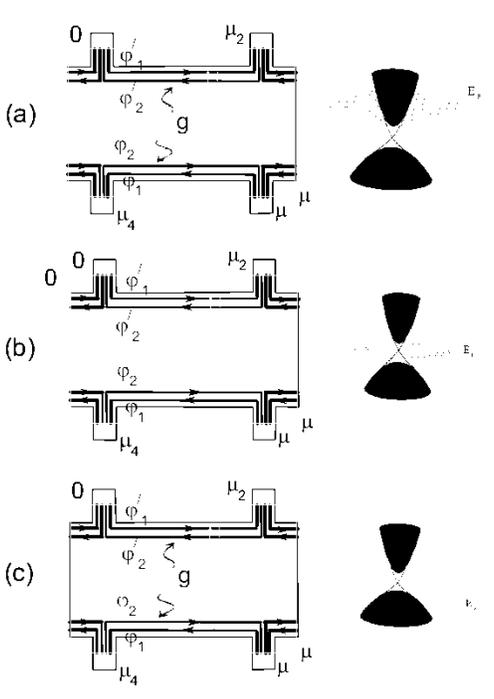}
\caption{
Schematics of edge state propagation for different gate voltages.
Parameter $\gamma$ describes  "edge to edge" elastic scattering,
parameter $g$- "edge to bulk" elastic scattering.
Insets—the energy spectrum for different Fermi energy positions in the region under local gate. }
\end{figure}

The local and nonlocal transport coefficients arise from the edge state contribution and the short-circuiting of the edge transport by the bulk contribution, the latter being more important away from the CNP. Figure 12b shows the comparison of the local conductance and nonlocal resistance demonstrating the interval of voltages, when the edge state transport is dominated. The nonlocal signal is zero when the Fermi level lies in the conduction or valence bands and far away from the CNP, and, when the classical model predicts vanishingly small nonlocal resistance. When the gate voltage is swept through the CNP the transitions between the edge states and the electron and hole bulk states continue which allows us to study the intermediate situation corresponding to an admixture of the edge and bulk contributions to the conductance.

\begin{figure}
\includegraphics[width=9cm]{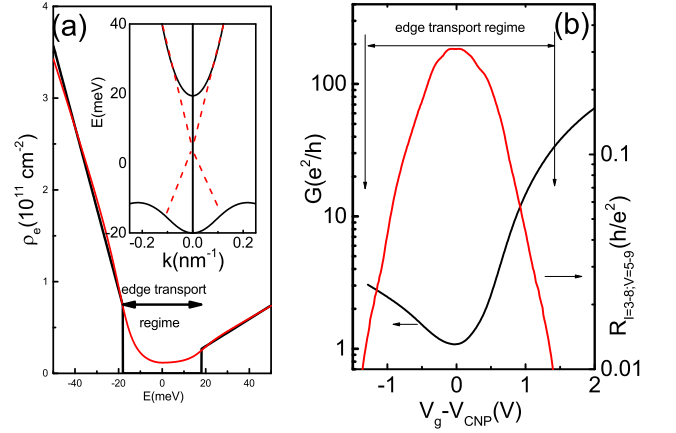}
\caption{(Color online)
a) The calculated density of states: black line-without disorder, red line- with disorder. Insert-schematic of the energy spectrum of topological insulator. (b)Conductance $G=1/R_{I=1-5;V=9-8}$ and nonlocal resistance $R_{I=3-8; V=2-9}$ as a function of the gate voltage, T=4.2K.}
\end{figure}

 The transport properties in the bulk can be described by the current-voltage relation\cite{abanin,Rahim}:

\begin{eqnarray}
\textbf{j}_{i}({\bf r}) =-\hat{\sigma}_{i}\nabla\psi_{i}({\bf r}),
\end{eqnarray}
\[
\hat{\sigma}_{i} = \left( {\begin{array}{cc}
 \sigma^{(i)}_{xx} & \sigma^{(i)}_{xy} \\
 \sigma^{(i)}_{yx} & \sigma^{(i)}_{xx} \\
 \end{array} } \right),
\]
where $i=1,2$ labels the states with different projections of the spin, $\psi_i$ are the electrochemical potential for electrons, and ${\bf r}=(x,y)$ is the 2D coordinate. Since we consider isotropic conduction, the non-diagonal part of the conductivity tensor appears only in a non-zero magnetic field. Assuming the components of the conductivity tensor to be coordinate-independent parameters, we can solve the problem by solving the Laplace equation for the potentials, $\nabla^{2}\psi_{i}({\bf r})=0$, because the charge conservation law and the continuity conditions require $\nabla\textbf{j}_{i}({\bf r})=0$. The solution to the Laplace equation is fully determined by the boundary conditions, which in our case are modified by the bulk-edge current leakage. In order to describe the transport in the presence of the edge states, we introduce two phenomenological constants $\gamma$ and $g$, which represent edge to edge and bulk to edge inverse scattering length, respectively. Then, the boundary conditions corresponding to a zero current normal to the boundary in the presence of a bulk-edge coupling are given by

\begin{eqnarray}
\textbf{n}\textbf{j}_{i} =g(\psi_{i}-\varphi_{i}),
\end{eqnarray}
where $\varphi_{i}$ are the local chemical potentials of the edge states, $\psi_{i}$ and $\textbf{j}_{i}$ are the potentials and currents at the boundary, and $\textbf{n}$ is a unit vector normal to the boundary. The edge state transport can be described by the continuity equations \cite{abanin, dolgopolov} taking into account the scattering between the edge and the bulk:

\begin{eqnarray}
\partial_{x}\varphi_{1}=\gamma(\varphi_{2}-\varphi_{1})+g(\psi_{1}-\varphi_{1}),\\
-\partial_{x}\varphi_{2}=\gamma(\varphi_{1}-\varphi_{2})+g(\psi_{2}-\varphi_{2}).
\end{eqnarray}

The general solution of this problem, therefore, includes the solution of a 2D Laplace equation for the bulk electrochemical potentials $\psi_{1,2}(x,y)$ together with Eqs. (11),(12),(13) describing the scattering between the edge states and between the edge and the bulk states. The current can be calculated from this solution as a sum of the contributions from the bulk and the edge states.

In nonlocal configurations the edge $+$ bulk model can be solved only numerically. We have performed self-consistent calculations to find the $\psi_{1,2}$ solution of the Laplace equation in two dimensions and the $\varphi_{1,2}$ solutions of equations 12 and 13 on the edge using the Hall bar geometry shown on the topo of the Fig. 5. The contacts are assumed to be thermal reservoirs, where a full mixing of electron spin states and bulk states occurs \cite{dolgopolov}. Note that, in contrast to the standard QHE, where the mixing of the edge states occurs within the metallic Ohmic contacts, in our samples the mixing will take place in the parts of the sample that lie outside of the metallic gate and contain 2D electron gas.

The equations for $\psi_{1,2}$ are discretized by the finite element method. The generalized Neumann boundary conditions, Eq. (11), are set in the regions outside the metal contacts. To solve the boundary value problem for a system of ordinary differential equations (12) and (13) we use a finite difference code that implements the 3-stage Lobatto IIIa formula. The boundary conditions inside the metal contacts are set to $\varphi_{1,2}=\psi_{1,2}$.

Both in local and non-local configurations the resistance is calculated as

\begin{eqnarray}
R_{xx} =V\;I_{tot}^{-1},\;I_{tot}=I_{edge}+I_{bulk}, \nonumber \\[0.6\baselineskip]
V =\frac{1}{2}\left( \varphi _{11}-\varphi _{11^{\prime }}+\varphi
_{21}-\varphi _{21^{\prime }}\right) ,
\end{eqnarray}

where $V$ is the potential difference at the voltage probes, $I_{tot}$ is the total current flowing between the current contacts, $\varphi _{i1}$ and $\varphi _{i1^{\prime} }$ are the potentials at the voltage probe locations. The edge and bulk currents for the local case (at an arbitrary point $x$ along the sample) are given by

\begin{eqnarray}
I_{edge}=\frac{e^2}{h} \left(\varphi_{1}-\varphi_{2}+\varphi_{2^{\prime }}-\varphi
_{1^{\prime }}\right), \nonumber \\
I_{bulk}=\sum_{i=1,2} \left[ \sigma_{xy}^{\left( i\right) }\left( \psi_{i}-\psi_{i^{\prime}} \right)
-\sigma_{xx}^{\left( i\right) } \int dy \frac{\partial \psi_{i}}{\partial x} \right],
\end{eqnarray}

where $\varphi_{i}$, $\psi_{i}$ and $\varphi_{i^{\prime }}$, $\psi_{i^{\prime}}$ are the potentials at the opposite (bottom and top, respectively) edges of the sample, and the integral is taken across the sample from bottom to top. For non-local case the currents are calculated from similar expressions:

\begin{eqnarray}
I_{edge} =\frac{e^2}{h} \left(\varphi_{1}-\varphi_{2}+\varphi_{2^{\prime }}-\varphi_{1^{\prime }}\right),
\nonumber \\
I_{bulk}=\sum_{i=1,2} \left[ \sigma_{yx}^{\left( i\right) }\left( \psi_{i}-\psi_{i^{\prime}} \right)
-\sigma_{xx}^{\left( i\right) } \int dx \frac{\partial \psi_{i}}{\partial y} \right],
\end{eqnarray}

where now $\varphi_{i}$, $\psi_{i}$ and $\varphi_{i^{\prime }}$, $\psi_{i^{\prime}}$ are the potentials at the opposite (left and right, respectively) edges of the current contact, and the integral is taken across this current contact from left to right. The conductivities are calculated as $\sigma_{xx}^{(1)}=\sigma_{xx}^{(2)}=e (\mu_n n+\mu_p p)/2$, where $\mu_n$ and $n$ ($\mu_p$ and $p$) are electron (hole) mobilities and densities. To find $n$ and $p$, the bulk densities of states for electrons and holes are represented by steps $D_{n,p} =4 \pi m_{n,p}/h^{2}$ with $m_{n,p}$ being the effective masses of electrons and holes. The energy gap separating electron and hole bands is $E_{g}=30$ meV. In addition, the sharp band edges are smeared according to the Gaussian law with the broadening energies $\Gamma_{n,p}= \hbar/\tau_{n,p}$, where $\tau_{n,p}=\mu_{n,p} m_{n,p}/e$. The following parameters have been used: $\mu_{n}=80000$ cm$^{2}$/V s, $\mu_{p}=5000$ cm$^{2}$/V s, $m_{n}=0.024~m_{0}$, $m_{p}=0.15~m_{0}$, where $m_0$ is the free electron mass. The comparison shown in figures 5 and 6 and is the representative behavior of a couple of local and nonlocal measurement configurations in representative device. The best agreement between the experiment and theory is reached for the value of phenomenological parameters $\gamma= 3.0 \mu m^{-1}$ and $g = 0.03 \mu m^{-1}$ . It is worth noting that the agreement between the calculations and the experimental data is much better than in the case of the Kirchoff's network model. Indeed all the local resistance values (6 possible configurations) also agree with calculations.

In the rest of the paper we would like to discuss the dependence of the nonlocal resistance on the density. Our model is much too simple to adequately describe the shape of the resistance peaks, shown in figure 8. The model reproduces the key feature of the nonlocal resistance, for example, a faster than in the case of a local resistance suppression of the peak away from CNP which is the result of a short-circuiting of the edge transport by the bulk contribution. However, we can not directly translate the energy dependence to the density dependence, because the Fermi energy does not vary linearly with $N_{s}$ in the bulk gap region. In the absence of disorder the Fermi level jumps from the conduction to the valence band, and a sharp resistance peak is expected, in contrast to the broad maximum observed in the experiment. The existence of the metallic puddles can be responsible for a smoother Fermi level displacement. For simplicity sake we can assume that the fraction of the metallic coverage of the sample is constant, which leads to a constant density of states inside of the bulk gap $\rho_{0}$. Comparing the energy and the density scales in figure 8 we obtain $\rho_{0}=5\times10^{10} cm^{-2}meV^{-1}$ which is close to the density of states of the electrons in the conduction band. We may assume that the metallic coverage $p<0.5$ is still below the percolation threshold and electrons are localized. Therefore, coexistence of the localized and delocalized electrons is needed for the description of the transport in a 2D TI. The localized electrons are responsible for a continuous transition of the Fermi level through the bulk gap, while delocalized carriers are responsible for a weak suppression of the nonlocal signal near the CNP and a its strong suppression away from the CNP. While our experiment offers an interesting outlook on the edge and bulk transport in a two dimensional TI, more experimental and theoretical work is required to understand the behavior of 2D electron system in such complex objects as disordered HgTe quantum wells.

\section{Concluding remarks}

Transport properties of 2D TIs have a number of specific features, related to the gapless spectrum of the helical edge states. For example, elastic backscattering is expected to be suppressed due to the time reversal symmetry. The mean free path and the mobility of electrons in the edge states of two-dimensional TIs can theoretically be very large. Considering the existing experimental results and analyzing the available theoretical work one can conclude that the experimental situation is still far from the ideal. 2D topological devices with long channels demonstrate a deviation of the conductance from the predicted $2e^{2}/h$ value. Note, however that the HgTe based 2D TI still remains attractive among the other alternatives, such as InAs/GaSb quantum wells \cite{knez,nichele,suzuki,qu}, and $WTe_{2}$ \cite{wu}. Observational studies can provide perspective on these issues, due to the possibility of improving the material quality in order to reduce potential fluctuations \cite{mikhailov} or fabricate devices in wider quantum well $d=14nm$ \cite{Olshanetsky}. The weak temperature dependence of the conductance at the CNP can be explained by mechanism, suggested in \cite{aseev}, related to the edge states coupling to large conducting puddles. The size and the density of the puddles can be significantly reduced by using cleaner starting materials, as has been done in the case of GaAs systems.

 Note that the observation of relatively high mean free path for the edge states in wide quantum well devices could be related to the advantages associated with the use of these structures \cite{Olshanetsky}. Indeed, the width of any quantum well is not uniform but fluctuates from point to point around its average value. These fluctuations are determined by the growth technology employed and is practically independent of the well width. The fluctuation of the well width results in a random potential in the bulk of the quantum well. However, the amplitude of that random potential would be much smaller in a wider well as it is proportional to $1/d^{3}$. From that point of view it is clear that a wider well is more advantageous for the observation of ballistic transport in 2D TI.

\section{Acknowledgements}

The financial support of this work by the Russian Science Foundation (Grant No.16-12-10041), FAPESP (Brazil), and CNPq (Brazil) is acknowledged.
E.B.O. acknowledges support of  RFBI Grant No. 18-02-00248a.

\end{document}